\documentclass{PoS}
\usepackage{amsmath}
\usepackage{amssymb,amsthm,amsfonts,bbm,dsfont,pifont}
\usepackage{slashed}
\usepackage{graphicx}

\newcommand{\PZ}{\text{Z}}
\newcommand{\PW}{\text{W}}
\newcommand{\ri}{\text{i}}

\title{Z+jet production at the LHC: Electroweak radiative corrections}
\ShortTitle{Electroweak corrections to Z+jet production at the LHC}

\author{Ansgar Denner\\
        Universit\"at W\"urzburg, Institut f\"ur Theoretische Physik und Astrophysik\\
        Am Hubland,  97074 W\"urzburg, Germany \\
        E-mail: \email{denner@physik.uni-wuerzburg.de}}

\author{Stefan Dittmaier\\
        Albert-Ludwigs-Universit\"at Freiburg, Physikalisches Institut, \\
        D-79104 Freiburg, Germany\\
        E-mail: \email{stefan.dittmaier@physik.uni-freiburg.de}}

      \author{\speaker{Tobias Kasprzik} \footnote{Preprint numbers:
          SFB/CPP-10-125, TTP10-49, FR-PHENO-2010-039, TTK-10-54.  The
          work of T.K.\ and A.M.\ is supported by the DFG
          Sonderforschungsbereich/Transregio 9 ``Computergest\"utzte
          Theoretische Teilchenphysik''.}\\
        Karlsruhe Institute of Technology (KIT),
        Institut f\"ur Theoretische Teilchenphysik, \\
        D-76128 Karlsruhe, Germany\\
        E-mail: \email{kasprzik@particle.uni-karlsruhe.de}}

      \author{Alexander M\"uck\\
        RWTH Aachen University, Institut f\"ur Theoretische Teilchenphysik und Kosmologie\\
        D-52056 Aachen, Germany\\
        E-mail: \email{mueck@physik.rwth-aachen.de}}

      \abstract{The investigation of weak bosons $V$ 
        ($V=\mathrm{W}^{\pm}$, $\mathrm{Z}$)
	produced with  or without associated hard QCD jets
        will be of great phenomenological interest at the LHC. 
	Owing to the large cross sections and the clean
	decay signatures of the vector bosons, weak-boson production can
        be used to monitor and calibrate the luminosity of the collider,
        to constrain the PDFs, or to calibrate the detector.	
	Moreover, the $\PZ$+jet(s) final state constitutes an 
	important background to a large variety of 
	signatures of physics beyond the Standard Model.	

        To match the excellent experimental accuracy that is expected at
        the LHC, we have worked out a theoretical 
          next-to-leading-order analysis of $V$+jet production at
        hadron colliders. The focus of this talk will be on new results
        on the full electroweak corrections to $\PZ(\to
          l^-l^+)$+jet production at the LHC. All off-shell effects are
        included in our approach, and the finite lifetime of the $\PZ$
        boson is consistently accounted for using the complex-mass
        scheme. In the following, we
	 briefly introduce
	the calculation and
        discuss selected phenomenological implications of our results.}

\FullConference{35th International Conference of High Energy Physics - ICHEP2010,\\
		July 22-28, 2010\\
		Paris France}

\begin{document}

\section{Introduction}
\noindent
The survey of Standard  Model weak-boson  production is an important
task in  the 
era  of LHC physics. The investigation of inclusive
$\PZ$-boson production is of special importance, since the production
cross section is comparably large, and the two charged leptons in the
final state allow for a precise event reconstruction, e.g.\ 
for a precise measurement of the invariant mass
$M_{ll}$ and the transverse momentum $p_{\text{T},ll}$ of the
intermediate boson. Therefore, such events are well suited to monitor
and to calibrate the luminosity of the collider, to determine the lepton
energy scale and the detector resolution, as well as to test
the linearity of
the detector response. Finally, the Drell--Yan process 
also plays an important role in a precise determination of the
$\PW$-boson mass and width.


At the LHC, $\PZ$ bosons will often be accompanied by one or more hard
QCD jets. On the one hand, such processes constitute a significant
background to various scenarios of physics beyond the Standard Model
that might be discovered at the LHC. On the other hand, the study of
$V$+jets ($V=\PW/\PZ$) events at the LHC may help us to gain a deeper
understanding of QCD and jet physics in general.

The next-to-leading-order (NLO) QCD corrections to $\PZ$+jet and
$\PZ$+2jets production at hadron colliders are known for a long
time~\cite{Giele:1993dj,Campbell:2002tg}. They are implemented in Monte
Carlo generators~\cite{Campbell:2002tg} and recently $\PZ$+jet
production has been matched with parton showers~\cite{Alioli:2010qp}. In
the past year, NLO QCD results for $V$+3jets and even $\PW$+4jets
production were presented~\cite{Berger:2010zx}.
However,
until now only the purely virtual weak corrections to on-shell $\PZ$+jet
production have been calculated for the LHC~\cite{Kuhn:2005az}, including  
next-to-leading-logarithmic and next-to-next-to-leading-logarithmic
approximations. 
 In Ref.~\cite{Kuhn:2005az}, the focus was on the
high-energy behaviour of the cross section and the dominating universal
high-energy logarithms.  
Complementary to those results, in this work we present the full NLO
electroweak (EW) corrections to off-shell $\PZ$+1jet production at
the LHC, taking into account the leptonic decay of the $\PZ$ boson to
allow for a realistic event definition.

\section{
Details of 
the calculation}
\noindent
 In this section we briefly introduce the setup of the calculation
which closely follows the setup explained in detail in 
Ref.~\cite{Denner:2009gj} in the context of $\PW$+jet production. For more 
process-specific details on $\PZ$+jet production we refer the reader 
to our forthcoming publication on the subject.

At tree level, three partonic channels contribute to the processes
$\text{pp/p}\bar{\text{p}}\to\PZ/\gamma+\text{jet}\to
l^-l^++\text{jet}$,
\begin{equation}
(i)\quad q\bar{q} \to \PZ/\gamma+\text{g}\,,\qquad\
(ii)\quad q\text{g} \to \PZ/\gamma+q\,, \qquad
(iii)\quad \text{g}\bar{q} \to \PZ/\gamma+\bar{q}\,,\nonumber
\end{equation} 
with $q =\text{u,d,c,s,b}$ denoting the active quarks. The QCD
parton in the final state (quark or gluon) will eventually be detected
as a hard jet in the hadronic calorimeter after hadronization. 

The finite lifetime of the $\PZ$ boson
is accounted for by including the corresponding decay width
$\Gamma_{\PZ}$ in the Z-boson propagator.
We work in the \emph{complex-mass scheme (CMS) for
  unstable particles}~\cite{Denner:2005es}, which enables a
consistent and gauge-invariant treatment of finite-lifetime effects in
one-loop calculations. In the CMS, the vector-boson masses $M_V$ are
consequently replaced by complex parameters, $M_V^2 \to \mu_V^2 = M_V^2
- \ri M_V \Gamma_V$, in the propagators \emph{and} in
the definition of all derived quantities, for example the weak mixing 
angle, 
i.e.\ $\cos\theta_{\text{w}}^2 \equiv \mu_{\PW}^2/\mu_{\PZ}^2$.

The computation of the full $\mathcal{O}(\alpha)$ corrections to
$\PZ$+jet production requires the calculation of real bremsstrahlung
corrections due to photon emission as well as the calculation of 
one-loop virtual corrections.  
Both real and virtual corrections give rise to so-called \emph{infrared
(IR) singularities} connected with soft and/or collinear photon
emission. These singularities are regularized either dimensionally or
alternatively via small lepton and quark
masses and an infinitesimal photon mass $\lambda$ and appear as $\ln
m_l,\,\ln m_q$, and $\ln \lambda$ terms in intermediate steps of the
calculation. In mass regularization
the $\ln \lambda$-dependence drops out after combining
virtual and real contributions, and residual $\ln m_q$-terms attributed
to initial-state photon radiation off quarks are absorbed in the
renormalized parton distribution functions (PDFs)
similar to a QCD factorization prescription. Since we
discard events with collinear parton--photon pairs in the final state if
the photon is sufficiently hard  to distinguish $\PZ$+jet from 
$\PZ$+photon production, the calculation is not collinear safe. Hence, it 
is necessary to introduce a
\emph{photon fragmentation
  function}~\cite{Buskulic:1995au,Denner:2009gj} to 
avoid
unphysical $\ln m_q$-terms in the physical cross section, which
 indicate that the collinear quark--photon physics cannot be understood
in a purely perturbative approach. 

In contrast to the quark masses, the lepton
masses have a well-defined physical meaning
and allow for the purely perturbative calculation of 
\emph{collinear-safe} and \emph{non-collinear-safe}
observables with respect to collinear lepton--photon splittings.
We consider event definitions with and without recombination of collinear
lepton--photon configurations in the electron and (bare) muon final state, 
respectively, observing corrections
enhanced by $\ln m_{\mu}$-terms in the latter case (see
Section~\ref{NumRes}).
We use an extended version~\cite{Dittmaier:2008md} of the 
dipole subtraction formalism which allows one to 
analytically extract the $\ln m_{\mu}$-terms 
for non-collinear-safe observables.



\section{Numerical results}\label{NumRes}

\noindent
In this section
\begin{figure}[t]
\begin{center}
\includegraphics[width=14.5cm]{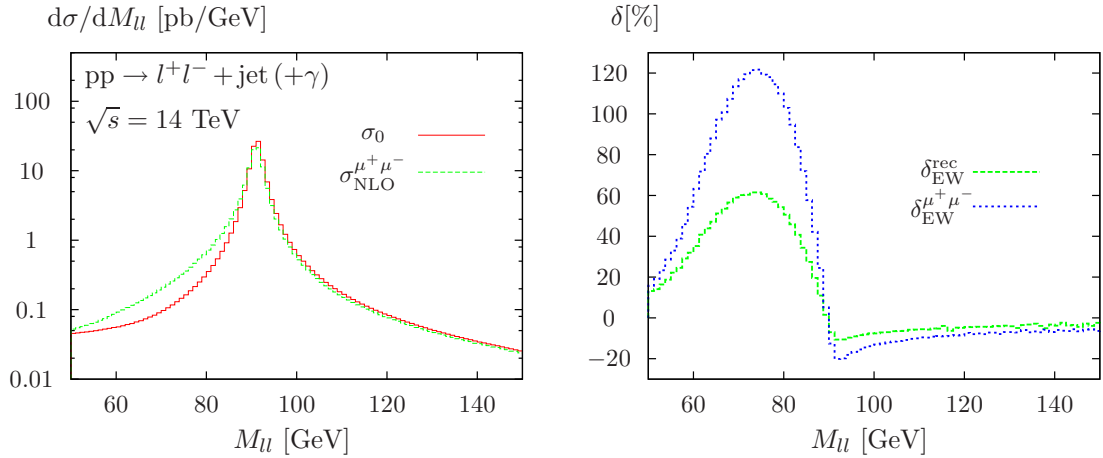}
\caption{\label{fi:mll} EW corrections (right) to the invariant-mass
  distribution of the lepton pair (left). Corrections for bare muons
  ($\delta_{\text{EW}}^{\mu^+\mu^-}$) and for recombination of collinear 
  lepton--photon pairs ($\delta_{\text{EW}}^{\text{rec}}$) are shown.}
\end{center}
\end{figure}
we discuss the distributions in the invariant
mass $M_{ll}$ and the transverse mass $M_{\text{T},ll}$ of the
final-state lepton pair, where we focus on the results for the LHC at
14\,TeV. 
The event-selection criteria applied in our calculation are similar to
the $\PW$+jet calculation which can 
be found in Chapter (3.2) of Ref.~\cite{Denner:2009gj}. 
For $\PZ$+jet production with two charged leptons in the final state, we
ask for a transverse momentum $p_{\text{T},l} > 25$ GeV  and a rapidity
$|y_{l}|<2.5$ for both leptons. Moreover, we require a minimal invariant
mass  $M_{ll} > 50$ GeV. 
\begin{figure}[t]
\begin{center}
\includegraphics[width=14.5cm]{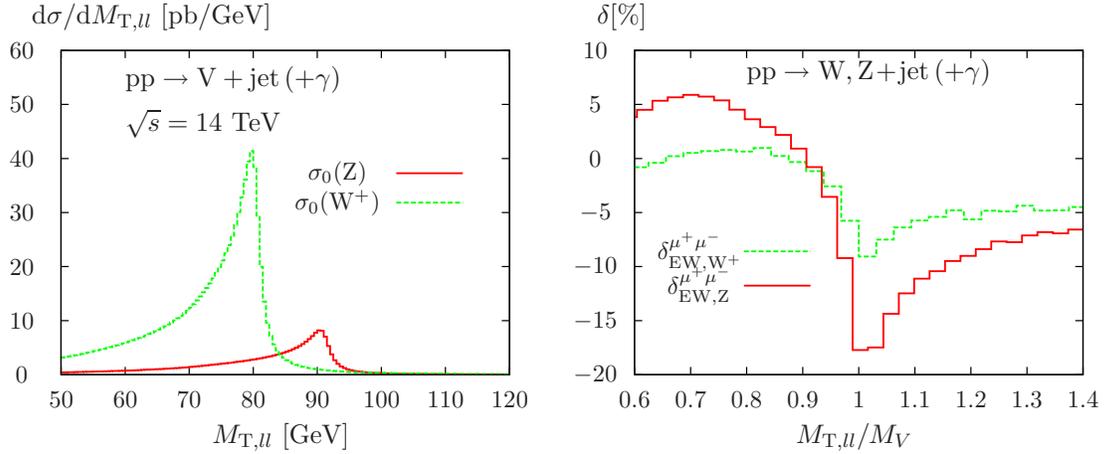}
\caption{\label{fi:mT} EW corrections (right) to the transverse-mass
  distribution of the lepton pair (left). Z+jet and W+jet
  production  are compared for bare muons.}
\end{center}
\end{figure}

Figure~\ref{fi:mll} shows the typical Breit--Wigner shape of the
$M_{ll}$ distribution at leading order (left) and the effect of the
relative EW corrections (right). We observe dramatic positive
corrections below the peak at $M_{\PZ}$ which are even larger than in
the single-Z case (see Fig.\ (12) in Ref.~\cite{Dittmaier:2009cr}), but
exhibit a similar qualitative behaviour. These huge effects can easily
be allocated to photon radiation off the final-state leptons, which
systematically shifts events to lower values of $M_{ll}$, where the
tree-level cross section is small. Of course, the
relative corrections $\delta_{\text{EW}}^{\text{rec}}$ for collinear-safe observables (electrons in the
final state)  are much smaller than the corrections
$\delta_{\text{EW}}^{\mu^+\mu^-}$ for bare muons, since in the
collinear-safe case electron and photon are recombined to a new
 (jet-like) quasi-particle that enters the cut procedure. Therefore,
the kinematics is not changed drastically in the collinear phase-space
region, where the matrix elements for photon emission are large.

The investigation of transverse-mass distributions allows one to
directly compare $\PW$- and $\PZ$-boson production, since
the analogue of
$M_{\text{T},ll}$ is also a well-defined observable for $\PW$
bosons. Concerning the LO cross section, the left-hand side of
Fig.~\ref{fi:mT} shows the Jacobian peak located at the vector-boson
mass and the rapid decrease for larger values of
$M_{\text{T},ll}$. Again, the EW radiative corrections (right) are
dominated by final-state photon emission; they induce positive
contributions below and negative contributions at the position of the
peak. Comparing the impact of the corresponding corrections for
$\PZ+\text{jet}$ and $\PW+\text{jet}$ production, respectively, we
observe that the effect is roughly a factor of two larger in the
$\PZ+\text{jet}$ case, because---contrary to the $\PW+\text{jet}$
situation---there are two charged leptons in the final state that may
emit a photon.

\section{Summary}
\noindent
We have calculated the full $\mathcal{O}(\alpha)$ corrections to
off-shell $\PZ+\text{jet}$ production with two charged leptons in the
final state for the LHC and the Tevatron, where the finite width of the
$\PZ$ boson is consistently accounted for using the complex-mass
scheme. Our approach is fully exclusive, allowing us to investigate any
differential cross sections and apply any event-selection cuts that are
of interest for experimentalists. The numerical analysis reveals
moderate corrections to the total cross section as expected, but we find
dramatic deviations in the line-shapes of fundamental leptonic
observables. The quantitative behaviour of the corrections turns out to
be significantly different compared to the single-$\PZ$ production
scenario, indicating that the indirect kinematic effects of the
additional hard jet on purely leptonic observables have to be
accounted for in a reliable analysis of LHC data.

\end{document}